\documentclass[10pt,journal,compsoc]{IEEEtran}

\usepackage{scalerel}
\usepackage{tikz}
\usetikzlibrary{svg.path}

\definecolor{orcidlogocol}{HTML}{A6CE39}
\tikzset{
  orcidlogo/.pic={
    \fill[orcidlogocol] svg{M256,128c0,70.7-57.3,128-128,128C57.3,256,0,198.7,0,128C0,57.3,57.3,0,128,0C198.7,0,256,57.3,256,128z};
    \fill[white] svg{M86.3,186.2H70.9V79.1h15.4v48.4V186.2z}
                 svg{M108.9,79.1h41.6c39.6,0,57,28.3,57,53.6c0,27.5-21.5,53.6-56.8,53.6h-41.8V79.1z M124.3,172.4h24.5c34.9,0,42.9-26.5,42.9-39.7c0-21.5-13.7-39.7-43.7-39.7h-23.7V172.4z}
                 svg{M88.7,56.8c0,5.5-4.5,10.1-10.1,10.1c-5.6,0-10.1-4.6-10.1-10.1c0-5.6,4.5-10.1,10.1-10.1C84.2,46.7,88.7,51.3,88.7,56.8z};
  }
}

\newcommand\orcidicon[1]{\href{https://orcid.org/#1}{\mbox{\scalerel*{
\begin{tikzpicture}[yscale=-1,transform shape]
\pic{orcidlogo};
\end{tikzpicture}
}{|}}}}

\usepackage{hyperref} 

\RequirePackage{amsmath}
\usepackage{bm}
\usepackage{bbm} 
\usepackage{graphicx}
\usepackage[long]{optidef} 
\usepackage{float} 
\usepackage{booktabs}
\usepackage{ltablex}
\usepackage{multirow}
\usepackage{stmaryrd}
\usepackage{color}

\DeclareMathOperator*{\argmax}{arg\,max} 

\ifCLASSOPTIONcompsoc
  \usepackage[nocompress]{cite}
\else
  \usepackage{cite}
\fi
\ifCLASSINFOpdf
\else
\fi
\begin{document}
%
\title{Continuous Prediction of Web User Visual Attention on short span Windows based on Gaze Data Analytics}
%
%
%
%

\author{Francisco Diaz-Guerra \orcidicon{0000-0002-2200-5705},
        Angel Jimenez-Molina \orcidicon{0000-0002-6866-6584}  
\IEEEcompsocitemizethanks{\IEEEcompsocthanksitem University of Chile, Department of Industrial Engineering. Beauchef 851, Santiago, Chile.\protect\\
E-mails: francisco.diaz.g@ug.uchile.cl, angeljim@uchile.cl}
\thanks{}}

%
%

\markboth{}%
{Shell \MakeLowercase{\textit{et al.}}: Bare Demo of IEEEtran.cls for Computer Society Journals}
%



\IEEEtitleabstractindextext{%
\begin{abstract}
The existing approaches to identify personalized salience zones of a Web page do not consider the dynamic behavior in time of the Web user's gaze or the alterations of its content. For this reason, this paper proposes the concept of visit intention, an indicator of the visual attention of a Web user in a certain period of time, short span time windows, in different areas of interest. This indicator gives information on the areas of interest of a website that will be visited by a user over a time window, without requiring to know the structure of the site in each window. Our approach leverages the population-level general gaze patterns and the user's visual kinetics. We show experimentally that it is possible to conduct such a prediction through multilabel classification models using a small number of users, obtaining an average area under curve of 84.3 \%, an average accuracy of 79 \%, and an individual area of interest accuracy of 77 \%. Furthermore, the user's visual kinetics features are consistently selected in every set of a cross-validation evaluation.
\end{abstract}

}

\maketitle

\IEEEdisplaynontitleabstractindextext

%
\IEEEpeerreviewmaketitle

\IEEEraisesectionheading{\section{Introduction}\label{sec:introduction}}

%
%
%
%

\IEEEPARstart{T}{he} behavior of the visual attention of the users is an active research area in the field of computer vision. Its applications allow the delivery of information of greater utility and/or complexity in areas that capture users' attention, and adapt a visual stimuli to users' gaze patterns in real time, among other advantages. In general, this research has tended to focus on images, and more complex visual stimuli have recently been integrated, such as videos \cite{bazzani2016recurrent}\cite{guo2008spatio}\cite{zhao2015fixation}\cite{zhai2006visual}\cite{fang2014video}, virtual reality environments \cite{sitzmann2018saliency}\cite{marmitt2002modeling} or egocentric videos \cite{huang2018predicting}\cite{zhang2017deep}. In addition, several studies of the behavior of visual attention in Web environments have been developed. Unfortunately, the latter are generally based on saliency maps computed on top of a static Web interface \cite{shen2014webpage}\cite{shen2015predicting}\cite{li2018webpage}\cite{li2016webpage}, or dynamic implementations that always have advance knowledge of the structure of the website \cite{xu2016spatio}. That is, the websites's structure is registered at all times, and static stimulus models are applied in periods of time when the website does not present structural changes. 


However, the main source of challenges for assessing the behavior of visual attention in current Web environments is related to the behavior of users themselves, such as the display of menus when clicking, the movement of the page given the lateral scroll, or the alteration of the design given the action of pop-ups or banners. Given that the interaction with a website presents constant changes, a prediction is sought that captures in a continuous fashion the areas of a user's visual attention without requiring the constant recording of the current structure of the stimulus, but knowing the structure that it presented in a past period of time. To address this challenge, this paper introduces the concept of ``visit intention'', defined as the probability that a user will fixate in a certain period to specific areas of interest (AOI) defined by means of groupings of website components. 

Visit intentions to AOI's are obtained through multilabel classifiers that leverage different features extracted from eye-tracking data. A first set of features is composed of gaze fixations in different periods over the components of AOI's. 

Since little information per user is available, our approach generates a population-level classification model to improve generalization and reduce the risk of overfitting that arises from the training of models with few samples. At the same time, a subset of these features embeds the information of the behavior of other users in that of a specific participant. That is, visit intention classification models integrate in a first instance the features calculated from a population-level gaze pattern, allowing predictors of individual behavior (with models of classification at the population level) to be trained in a robust way. 

Another set of features, such as position, velocity and acceleration measurements in the X and Y coordinates, includes what we call the user's visual kinetics. We hypothesize that these features allow us to capture tendencies of the gaze movement at the beginning of a period, which would allow knowing the future eye movement patterns.

This paper thus attempts to answer the following research question:

\begin{itemize}
\item RQ:\textit{ To what extent is it possible to accurately classify in real time a user's visit intention in a certain period of time to specific AOI's of a website, leveraging population-level general gaze patterns and a user's particular gaze data?}
\end{itemize}

To answer this research question, an experiment was conducted in which 51 users browsed a website with dynamic components in front of an eye-tracking device. This paper shows that by using population-level gaze patterns of a small number of users and applying diverse classification methods and feature selection techniques, a user's visit intention to AOI's in a period of time can be classified as proposed in RQ (average AUC = 0.843, average ACC = 0.79). Furthermore, a user's visual kinetics features are consistently selected in every set of a cross-validation, which confirms our hypothesis. 

This paper is organized as follows. Section \ref{RelatedWork} presents the related work. Section \ref{Methods} defines key concepts, delivers a formulation of the problem to be addressed and explains the apparatus used for the experimentation carried out to answer the research question. Section \ref{PredictionMethodology} explains the proposed prediction methodology. Section \ref{ExperimentalResult} shows an experimental evaluation by applying the proposed methodology to a real dataset of Web users' gaze data. Section \ref{Discussion} presents the discussions of the work carried out, while the paper is concluded in Section \ref{Conclusions}.

 

\section{Literature Review} \label{RelatedWork}

Visual attention is a mechanism that filters relevant areas of a stimulus from the rest of redundant information. Knowledge of a user's attention can be useful in applications such as coding and data transmission, improvement in recommendation and information delivery systems, and performance evaluation in different visual stimuli, among others.

In the literature, attention is categorized into two functions: bottom-up attention and top-down attention. The first corresponds to the selection of zones of a visual stimulus based on its most salient inherent characteristics in relation to the background. This function of attention operates with input in a crude, involuntary way. The second integrates knowledge of the visual scene, goals or objectives of the user with respect to the stimulus. Moreover, it implements longer-term cognitive strategies on the part of the person \cite{katsuki2014bottom}\cite{rutishauser2004bottom}\cite{connor2004visual}.

Top-down attention models are less proposed in the literature than the bottom-up case, since their functionality is influenced by factors external to the stimulus. In addition, calculated responses vary from model to model, i.e., from task to task, making it difficult to generalize a stimulus model. On the other hand, different models of bottom-up visual attention have been proposed. In this work, bottom-up models are studied.

\subsection{Attention Models in Static Stimulus}
The focus of classic studies on visual attention corresponds to static visual stimuli, that is, stimuli that do not present an alteration in the time of exposure in front of the user. In this field, there are multiple studies about visual attention when making ``free viewing'' on images.  Of these, there are mainly three approaches for the study of visual attention: saliency map generation, scanpaths models and the saccade models.

Saliency maps are topographic representations of the visual prominence of a stimulus. That is, they make it easier to understand areas of greater or lesser importance in the selection of points of attention. There are multiple studies in which saliency maps are applied to images. In \cite {kruthiventi2016saliency}, a deep convolutional neural network is proposed that is capable of predicting fixations and segmenting objects in different images. In \cite {johnson2017study}, transformations are made from features present in different natural images to maps of salience, searching for a smaller number of characteristics to use. In \cite {tang2017prediction}, the images are convolved with a template and postprocessed to deliver the result. In general, there is a large database of images and different models developed in the state-of-the art approaches \cite {bylinskii2015saliency}, as well as studies about their performance metrics \cite {bylinskii2018different} \cite {le2013methods}.

Scanpaths are attempts to predict sequences of users' fixations within a stimulus. While many papers study scanpath generation and saliency map extraction independently, research focusing on the fusion of these approaches is limited. However, some works focus on the identification of scanpaths by using sampling methods on top of saliency maps \cite{shen2014webpage}\cite{shen2015predicting}.

Saccadic models represent another option. In \cite{wang2011simulating}, predictions of exploration routes are made in natural images. In \cite{meur2017computational}, the saccadic model considers age ranges of users. This model integrates the interaction between the way in which users observe the visual information and the mental state of each user.

It has been demonstrated that static models present good results; however, in this work, we seek to study the visual attention of users in dynamic environments, which in general are more complex since they consider ``time'' as a new variable. 

\subsection{Attention Models in Dynamic Stimulus}
There are models developed for bottom-up attention on dynamic stimuli, that is, environments that present movement or that change their structure when interacting in a certain way with the person. For example, a video presents alterations (frame-to-frame movement), but it does not depend on the interaction. On the other hand, a web page presents dynamism in its structure or content depending on the user's interaction.

In \cite {bazzani2016recurrent}, predictions of saliency maps are made in videos, frame-by-frame. For this purpose, recurrent network methods with mixed density are used. In this study, the visual stimulus is known over time, so it could be broken down into a series of static images. However, the advantage of its approach, in which the saliency map for each frame of the video is generated by the information from the previous frames, is that they integrate information from the user interaction.

Saliency maps have also been applied to stimuli in web environments. In \cite {zheng2018task}, a saliency map is calculated at the task level. However, most studies have tended to focus on static calculations \cite {shen2014webpage} \cite {shen2015predicting} \cite {li2018webpage} \cite {li2016webpage}. That is, they do not have the temporal component of the user's interaction with the website. Finally, there is a dynamic visual attention study \cite {xu2016spatio} in which the site is assumed to be known at all times, and at the same time, it is assumed to know the duration of each subtask conducted by the user.

Although these methods correspond to a dynamic stimulus and contemplate information of user interaction, they generally focus on knowing the stimulus at all times, either by its nature independent of the user or by predicting the duration of subtasks (predicting the structure of the stimulus in certain periods of time). Our approach seeks to generate predictions directly from visual data, without the need for prediction of changes in the stimulus or its structure control at all times, seeking to accelerate the processing times of the model in execution. 

\subsection{Attention Models in Egocentric Vision} \label{egocentric}
A recent area of study is called egocentric vision. This is the analysis of attention in videos captured with wearable cameras on a user such that the visual fields during the execution of different tasks can be determined. In general, cameras are used on the head or chest of the user, and in addition to recording the environment in which it is carried out, the eye activity is recorded by an eye tracker. 

A recent study on gaze prediction in these environments is \cite{huang2018predicting}. Here, the authors look for the generation of saliency maps in future frames through frame prediction, a similar approach to the one used in \cite{zhang2017deep}. Although these environments are dynamic and depend on user interaction, the possibilities for interaction are so broad that traditional bottom-up models are insufficient for visual prediction \cite{yamada2010can}. Other models developed do not achieve generalization because they have specific cases for each task \cite{yamada2011attention} \cite{li2013learning}.

As in the case of the dynamic environments described above, the models consider the prediction of a visual stimulus structure or future sub-tasks for the estimation of visual attention. This is because the stimuli in this area are too complex, escaping our framework. In particular, what we are looking for is to study a dynamic stimulus that moves within well-defined structures.

To the best of our knowledge, this is the first work to study attention transition between zones of a dynamic visual stimulus that changes according to the user's behavior, directly from user's visual information, without predicting changes in the stimulus structure.

\section{Material and Methods} \label{Methods}

\subsection{Problem Definition} \label{ProblemDefinition}

This paper studies the visual attention of a user on a dynamic stimulus, within a bottom-up attention environment, that is, where attention is guided by inherent properties of the stimulus, and not by factors internal to the user such as previous knowledge of the stimulus or specific tasks to be carried out (top-down attention). To study the above topic, a two-dimensional stimulus is created and is exposed to different users, and each interaction is recorded with Eye Tracker instrumentation. 

User visual behavior is represented by the recording of gaze data. In this way, each point of gaze $x_i$ is represented by a vector of three components $x_i = [x_i, y_i, T_i]$, where the first two correspond to the horizontal and vertical coordinates of each record over a two-dimensional stimulus, and the third corresponds to the time in interaction. In the following, a \textbf{signal} is referred to as the visual record of a complete interaction.

The Eye Tracker records a user's visual signals. This gives us information of gaze, being able to identify fixations and saccades. A \textbf{fixation} is the state in which the eyes remain static, with some micro-movements, for a period of time (100 - 300 milliseconds), for example, over a word when reading. They are identified as sets of gaze data, and their position is considered as the place where the user keeps their attention. A \textbf{saccade} corresponds to ballistic movements that have a duration between 30 and 80 milliseconds, made between one fixation and another. This is not necessarily done in a straight line.

The development of this work is based on the segmentation of stimulus into \textbf{areas of interest (AOI)}, that is, regions where the researcher is looking for user behavior.
The space between the AOI's is called \textbf{whitespace}, which is not necessarily free of stimulus elements. 

For the AOI definition, there is an approach where the stimulus is divided into equal size regions, called segmentation in grid.
 On the other hand, AOI's can be defined in a semantic way, such that common meanings are used between zones assigned to the same AOI. There may be differences in shape and/or size with respect to other AOI's. We use the latter approach in this work given that it is expected to obtain information about the visual attention of the user in certain areas with a dynamic nature.

Therefore, an important factor in the problem is that for each AOI definition, it is necessary to consider grouping the zones of the stimulus that present the same information for the user. For example, grouping two news items within a newspaper allows the AOI to be characterized as news information. However, grouping an advertisement with one news item does not allow the AOI to be given a characteristic meaning.

This paper considers the visual attention of users to the different AOI's defined within a dynamic stimulus. For this, the concept of \textbf{``visit intention''} is defined, by which it is evaluated in advance such that the AOI visual stops by a user will be verified. In this way, given the visual stimulus, the AOI's $\lbrace A_{j} |\ j=1...n \rbrace$ and a period of time  $v^{(t)} = (T_o^{(t)}, T_f^{(t)})$ are given, where the values of the tuple correspond to the time limits of the period, which are indexed by $\lbrace t=1,2,... \rbrace$, and the visit intention $Iv(A_{j},v^{(t)}) \in \lbrace 0, 1\rbrace$ is defined as a binary variable that indicates whether or not there is a user decision to visit $A_j$ in that period. The decision to visit an AOI will be measured as the existence of visual fixations of the user within that stimulus zone. The visit intention for $n$ AOI's in a period time is a vector $IV(v^{(t)})= [Iv(1,v^{(t)}),...,Iv(n,v^{(t)})]$.

It is assumed that the vector $IV(v^{(t)})$ is characterized at the beginning of the time period $v^{(t)}$, although in reality, the process of choosing it can take place during all the time before the beginning of the window $v^{(t)}$. What we are looking for in the following sections is the prediction of this vector for different time windows $v^{(t)}$, given the information of the user's behavior in the past time (periods $v^{(s)},\ s= 1, \dots, t-1$).

What follows from this section describes the experiment developed to study the bottom-up visual attention of different users on a dynamic visual stimulus based on the estimation of the values of ``visit intention'' in different time periods.

\subsection{Participants}

The experiment is conducted with 51 participants based on a convenience sample. Of these participants, 34 are men and 17 are women, between the ages of 19-35. A large percentage of the participants (88\%) are students, and the occupations of the rest include a research assistant, a participant with a bachelor's degree in Hispanic literature, and four engineers.

The participants signed an informed consent form stating that they allow the use of the data collected and meet the following requirements:

\begin{itemize}
    \item They are healthy individuals who do not present with diseases that would harm the results of the experiment, they are not under pharmacological treatment and they do not have a history of neurological or psychiatric diseases.
    \item They do not present with the use of medications or drugs within the past 24 hours.
    \item They present with good vision or corrected vision.
\end{itemize}

\subsection{Apparatus}\label{Apparatus}

\subsubsection{Instrumentation:} 

The Eye Tracker Tobii T120 (Tobii Technology AB, Sweden) is used, which allows tracking of the subject's gaze during the test.
This setup corresponds to a screen of 17 inches, with a resolution of 1280 x 1024 px, where two infrared diodes generate patterns of reflection on the corneas of the user's eyes; the three-dimensional position of each eye is calculated, and therefore, his or her gaze on the screen is determined. 
At the same time, the pupil sizes of both eyes are recorded over time. The Tobii Studio software allows the adjustment of the user's position during calibration.

The system has a sampling rate of 120 Hz, a typical accuracy of 0.5 Hz, a typical drift of 0.1 degrees, a typical spatial resolution of 0.3 degrees and a typical head movement error of 0.2 degrees.

\subsubsection{Web Page:} 

To carry out the study, a dynamic stimulus corresponding to a web page is used. To achieve the dynamic character, the static and complete website version cannot be seen entirely on the screen, requiring the use of a scrolling bar by the user. At the same time, dynamic elements within the page, such as drop-down menus or moving elements, are considered.

\begin{figure}
\centering
\includegraphics[width=0.45\textwidth]{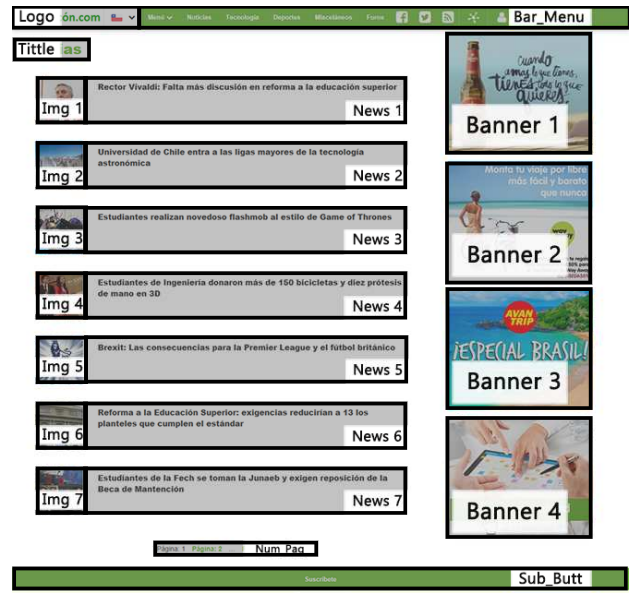}
\caption{Static components: Web page components.}
\label{WebPage}
\end{figure}

\begin{figure}
\centering
\includegraphics[width=0.45\textwidth]{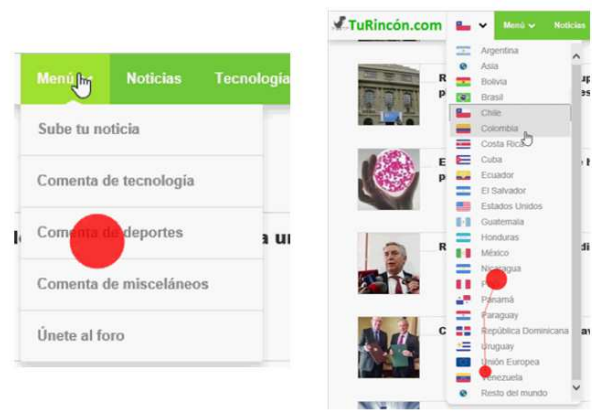}
\caption{Dynamic component: Page drop-down menu.} \label{MenusDesplegables}
\end{figure}

\begin{figure}
\centering
\includegraphics[]{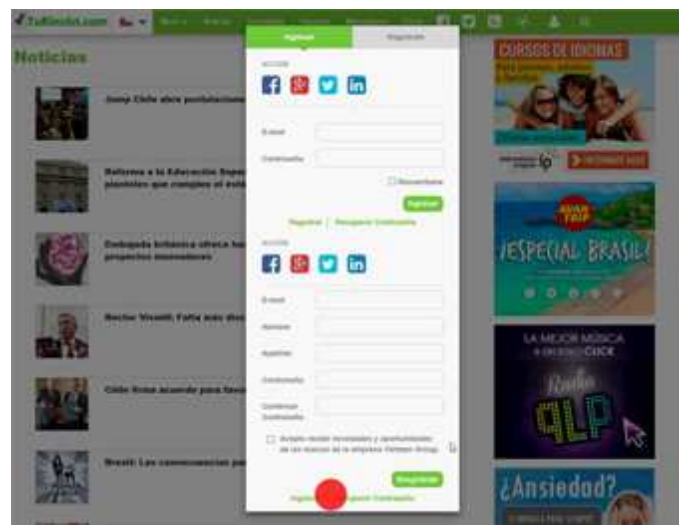}
\caption{Dynamic component: User registration pop-up.} \label{RegistroUsuarios}
\end{figure}

A web page has different \textbf{components}, that is, the sections of the page that represent an element in its own functions and actions. Web page components can be seen in figure \ref{WebPage} with different colors, and correspond to the following: ``Logo''; ``Bar\_Menu''; ``Title''; ``Img'' and ``New'' for each of the seven news items; ``Banner'', for each of the four advertisements; ``Num\_Pag'' (corresponding to the pages number button); and ``Sub\_Butt'' (corresponding to the user's subscription button). In addition, there are dynamic components, that is, they appear according to the user's interaction with the website. These are the country selection bar, the menu bar presented in figure \ref{MenusDesplegables} and the user registration pop-up window in figure \ref{RegistroUsuarios}.

\subsection{Task Design}

The objective of the experiment is to predict the attention zones of a user's gaze during the exposure of a dynamic stimulus in a bottom-up attention task. To perform this, a fictitious website is chosen, created specifically for the experiment, which is run locally to avoid possible delays in connection to the internet. The web page is detailed in section \ref{Apparatus}.

Three websites are presented to each participant. Each considers the same structure in the elements of the page, but the content varies. Advertisements and news headers appear randomly and without repetition, out of a total of 12 available advertisements and 21 news items. This approach avoids the experience effect of a user with the stimulus and, at the same time, creates a dynamic stimulus with different content for each user. The aim is that users navigate freely on the website and that this, together with the lack of previous experience of the user with the page, allows the study of bottom-up attention.

The content that is chosen, both for news and announcements, is related to student life. Topics include travel, courses, beer, education, music, technology and miscellaneous.

For the best analysis of the data, the control of variables such as luminosity and the user's movement is required, such that an isolated space is necessary to carry out the study. For eye tracking, direct sunlight, which affects the quality of the measurements, should be avoided. In addition, there should be no light from a top source. To avoid these effects of light, the laboratory is isolated from external light with black-out curtains. In addition, measures are taken to avoid all kinds of external interruptions during the experiment.

\subsection{Experimental Procedure}

During the experiment, only the person in charge of the test and the participant are in the experimental room. The latter is provided with an explanation of the experiment and asked to read and sign the informed consent form. 

The person sits in front of the Eye Tracker screen in a manner similar to navigating on a desktop computer, allowing the use of the mouse. The participant is asked to maintain a fixed posture, without moving his or her head, to avoid data loss.

Each person is prompted to freely browse the web page for as long as he or she wants and to indicate when to finish.

Prior to the tests, each user is subjected to a relaxation stage consisting of the visualization of three videos of four minutes each, consisting of beautiful landscapes with background instrumental music. The second part of this stage asks the participant to take a deep breath for one minute with closed eyes and soft instrumental music in the background. This stage aims to eliminate the Hawthorne effect (behavior modification as a consequence of knowing they are being studied) and physiological effects. In addition, this stage allows the baseline to be acquired for later treatment of the signals. Then, the page is presented randomly to each participant.

\subsection{Descriptive Analysis}

\begin{table}
  \centering
  \caption{Statistical Analysis}
    \begin{tabular}{cc}
    \hline
    Number of records & 153 \\
    \hline
    Recording time (s) & $t_{min}=16.5$ - $t_{max}=399$  \\
          & $\bar{t}=78.8$ - $\sigma_{t}=51.9$ \\
    \hline
    Number of saccades & $s_{min}=62$- $s_{max}=1129$ \\
          & $\bar{s}=262$ - $\sigma_{s}=165$ \\
    \hline
    Fixation per record & $F_{min}=62$ - $F_{max}=1130$ \\
          & $\bar{F}=262$ - $\sigma_{F}=164.6$ \\
    \hline
    Fixing duration per record (s) & $F_{min} =0.008$ - $F_{max} =3.273$ \\
          & $\bar{F}=0.22$ \\
    \hline
    \end{tabular}%
  \label{tab:statistics}%
\end{table}%

A total of 153 signals are extracted from 51 volunteers for the signals of the gaze behavior and event registration (click, display of dynamic elements, scroll, etc.). Fixations with duration less than 100 ms are filtered from the model. In table \ref{tab:statistics}, there is an exploratory analysis of the acquired signals.

\section{Prediction Methodology} \label{PredictionMethodology}

\begin{figure*}
\centering
\includegraphics[width=0.9\textwidth]{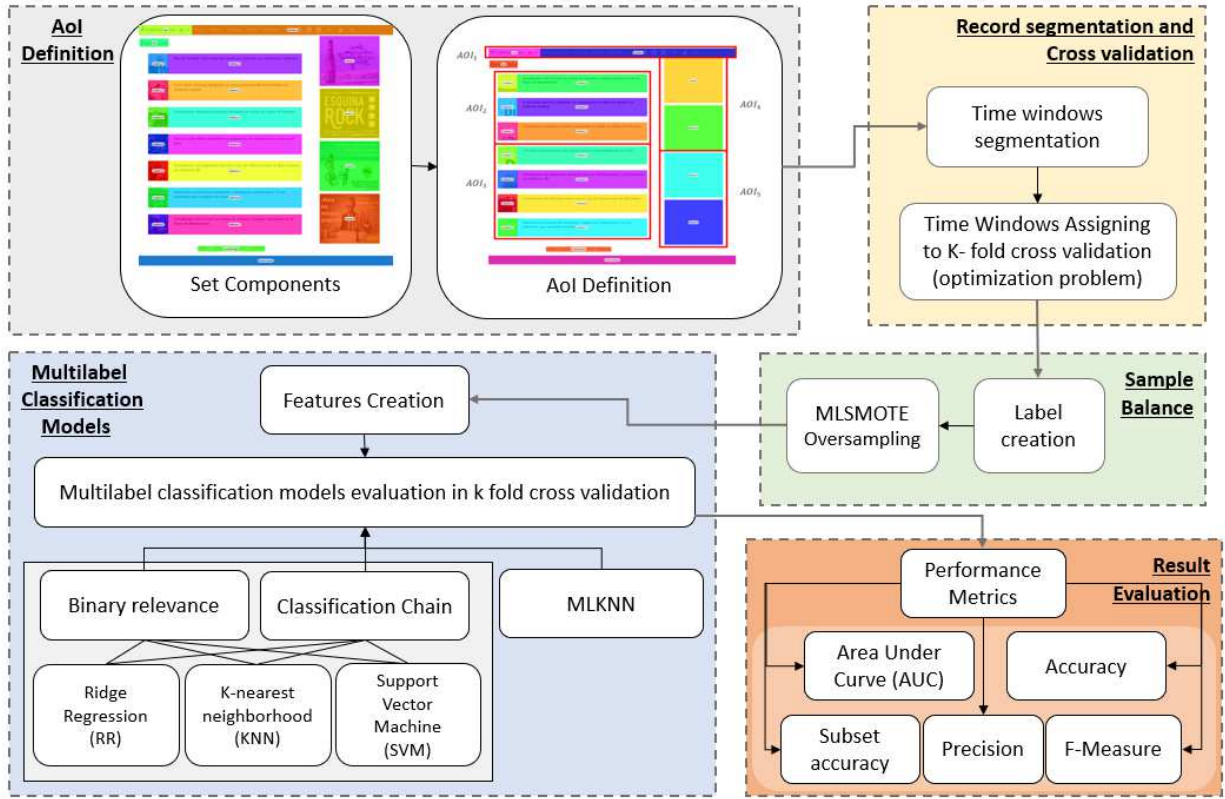}
\caption{Prediction Methodology} \label{Methodology_image}
\end{figure*}

The figure \ref{Methodology_image} shows the methodology for the construction of predictive models of visit intention in limited periods of time. First, components of the web page are established, and then, AOI's are defined. Afterwards, signal records are divided into time windows of a specific length by solving a linear optimization problem. The time windows are then optimally assigned to training and test sets for cross-validation. 
Labels for each time window are created, and a oversampling balance method is used for each train-test set to avoid overtraining. For each time window, it is possible to create a set of features used for classification. Different methods are tested, and finally, the results are evaluated.

\subsection{AOI Definition Criteria}

Given the approach of the problem, each defined AOI must have a characteristic meaning; therefore, components of similar semantic meaning were grouped into larger areas. At the same time, the final size of each defined area is considered so that it can be fully visible on the screen by scrolling the page in the corresponding dimensions.

\begin{figure}
\centering
\includegraphics[width=0.45\textwidth]{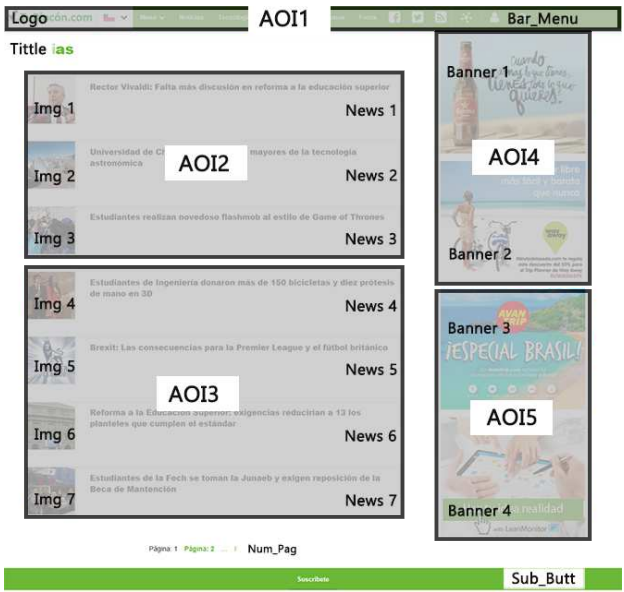}
\caption{AOI definitions for the web page.} \label{separacionAOI}
\end{figure}

In this way, the web page is separated into five semantic areas shown in figure \ref{separacionAOI}. A sixth AOI is defined, corresponding to the registration menu that is displayed when a user clicks on the button intended for this functionality, located in the upper navigation bar (see figure \ref{RegistroUsuarios}). 

AOI 1 is dynamic, as the menu bar moves together with the screen movements given by the scroll. The AOI's 2 and 3 are groupings of outstanding news, and the division into three and four news groups is chosen since the resolution of the screen allows these AOI's to be completely displayed when performing a lateral scroll.
 Following the same logic, AOI's 4 and 5 are generated corresponding to advertising banners within the web page.

AOI 6 has a dynamic nature and is activated by user interaction within the site. When activated, the ocular fixations for the rest of the AOIs located below are deactivated.

Some components of the page are not included in any AOI, as is the case for the subscription bar, the number of pages and the title of news; these are considered to be whitespace.

Thus, each defined AOI meets the conditions of the problem, where AOI 1 acquires the meaning of the navigation menu, areas 2 and 3 refer to news, areas 4 and 5 to advertising and AOI 6 to the user registration menu.

\subsection{Time Window Segmentation and Cross-Validation}

The models developed simulate the continuous acquisition of gaze data over time. To evaluate this acquisition, the gaze signal is segmented into time intervals called windows. These windows have a long $\tau$, given as a parameter. Thus, a long signal $T$ will have $t = 1, \dots, \lfloor T/\tau \rfloor$ time windows $v^{(t)}$.

To achieve external validity of the prediction models, it is necessary to generate training and test datasets that maintain independence from each other. To achieve this generalization, it is considered that both sets cannot share information from the same user. Thus, using the K-fold cross-validation method, a partition per user is performed. In turn, to select which users belong to each of the K-folds, an optimal allocation is performed that balances the number of windows in each fold. Since each user's total browsing time is different, the number of users in each partition varies. 

The assignment is made with a mixed integer linear programming model given by equations (\ref{eq:Fobjetivo}) to (\ref{eq:NatY}).Sets $U = 1, \dots, N$ and $KF = 1, \dots, K$ represent the number of users and fold to be used in cross validation, respectively. Variables $\lbrace y_{u,k}| u \in U; k \in KF \rbrace$ are defined with the value one if the user  $u$ is assigned to fold $k$; otherwise, they are zero. The variables $\lbrace x_{k} |k \in KF \rbrace $ represent the sum of windows in fold $k$, given the assignment of users. Thus, each user is assigned to a fold, minimizing the difference between the total sum of windows in each fold $k$ and the average number of windows in all folds. In the best case, all folds will have a number of windows equal to the average $\bar{x}$.

The restriction (\ref{eq:UnFoldXusuario}) indicates that each user can be assigned to only one of the K-folds. The restriction (\ref{eq:DefX}) defines the variables $x_{i}$, each representing the weighted sum between the assigned users and the total number of windows it has in the interactions ($P_{u}$). The restriction (\ref{eq:DefXbarra}) defines the arithmetic mean of the number of windows in the folds.

\begin{mini!}|l|[2]                   
    {}                                
    {\sum_{k=1}^{K}|x_{k}-\bar{x}| \label{eq:Fobjetivo}}   
    {\label{eq:ParticionK-Fold}}      
    {}                                
    \addConstraint{\sum_{k=1}^{K}y_{u,k}}{=1 ,}{u \in U \label{eq:UnFoldXusuario}}
    \addConstraint{x_{k}}{=\sum_{u=1}^{N}y_{u,k}*P_{u} , \qquad}{k \in KF \label{eq:DefX}}
    \addConstraint{\bar{x}}{=\frac{\sum_{k=1}^{K}x_{k}}{K} \label{eq:DefXbarra}}
    \addConstraint{x_{k}}{ \geq 0, }{k \in KF \label{eq:NatX}}
    \addConstraint{y_{u,k}}{\in \lbrace 0, 1\rbrace, }{u \in U; \ k \in KF \label{eq:NatY}}
\end{mini!}
\\

To solve (\ref{eq:ParticionK-Fold}), we add new variables $U_{i},$ $i=1,\ldots,K$ that represent $|x_{i}-\bar{x}|$. The target function is changed, and constraints (\ref{U1}) and (\ref{U2}) are added. This is how problem (\ref{eq:ParticionK-FoldRelax}) is defined.

\begin{mini!}|l|[2]                   
    {}                                
    {\sum_{k=1}^{K}U_{k}}   
    {\label{eq:ParticionK-FoldRelax}}     
    {}                                
    \addConstraint{U_{k}}{\geq x_{k}-\bar{x} ,}{k \in KF \label{U1}}
    \addConstraint{U_{k}}{\geq -(x_{k}-\bar{x}), \qquad}{k\in KF \label{U2}}
    \addConstraint{U_{k}}{ \in {\rm I\!R}}
    \addConstraint{\lbrace (\ref{eq:UnFoldXusuario}), (\ref{eq:DefX}), (\ref{eq:DefXbarra}), (\ref{eq:NatX}), (\ref{eq:NatY}) \rbrace \qquad}
\end{mini!}

\subsection{Sample Balance} \label{Balance}

In multilabel classification problems, it is common that label frequencies are not the same, with some that are active more frequently than others. In this case, since labels are defined by combinations of AOIs, vectors will exist with few combinations, as in the cases in which the dynamic areas of interest are activated, such as the pop-up windows, which are carried out to a lesser extent than the activation of the static AOIs.

The MLSMOTE ~\cite{MLSmote} resampling method is used to solve the sample imbalance problem. This method uses the minority class labels as seeds for the generation of new instances. For each minority class label, it searches for the nearest neighbors by obtaining the characteristics of the new samples using interpolation techniques.
Therefore, the new instances are synthetic rather than clones of other data. 

\subsection{Features Creation for Multilabel Classification}

For multilabel classification, a set of independent features is generated, which are processed from gaze data signals and navigation history. Indexes $j =1 , \dots, n$ correspond to the defined AOI's and $c = 1, \dots, N$ to the components present in the web page.

The features related to the \textbf{ocular fixations} carried out in each component and AOI are extracted, which are defined in table \ref{variables}. These features attempt to capture attention in certain areas given the history of previously visited areas. 

\begin{table*}
    \centering
    \caption{Predictive ocular fixation model features}
    \begin{tabular}{c|p{14 cm}}
        \hline
        \multicolumn{1}{c}{\textbf{Feature}} & \multicolumn{1}{c}{\textbf{Definition}}\\
        \hline
        \\
        $\lbrace AOI\_tslv \rbrace_{j}$ & The ``time since last visit'' features for each AOI are calculated as the difference between the start time of the current window and the time of the last time that user make a fixation in the j-th AOI ($\lbrace AOI \rbrace_{j} = 1$). Value 0 in case of no previous visit.\\
        \\
        \hline
        \\
        $\lbrace AOI\_r1 \rbrace_{j}$ & \multirow{3}{=}{\setlength\parskip{\baselineskip}%
        Independent time series features. $rq$ correspond to 1 if the user makes a fixation in the q-th window previous \ to the current one without temporal accumulation, 0 if not. That is, if $\lbrace AOI \rbrace_{j}=1$ for a window $v^{(t)}$, then $\lbrace AOI\_{rT} \rbrace_{j}=1$ for the window $(v^{t+T})$.}\\
        $\lbrace AOI\_r2 \rbrace_{j}$ & \\
        $\lbrace AOI\_r3 \rbrace_{j}$ & \\
        \\
        \hline
        \\
        $\lbrace AOI\_end\_r1 \rbrace_{j}$ & Binary feature. Takes the value $1$ if the previous window ends with a fixation in the j-th AOI and $0$ if not. This feature captures the fact that fixations can be interrupted by transitions between windows.\\
        \\
        \hline
        \\
        $End\_r1$ &  Corresponds to an AOI indicator in which the previous window ends. It is calculated as:\\
         & \multicolumn{1}{c}{$End\_r1=\sum_{j=1}^{n} j*AOI\_end\_r1_{j}$}\\
         \\
        \hline
        \\
        $\lbrace [Component]\_his \rbrace_{c}$ & The ``history'' feature for each component. Takes the value $1$ if user has made a fixation on the component in the past and $0$ if not.\\
        \\
        \hline
        \\
        $\lbrace [Component]\_end \rbrace_{c}$ & This feature indicates for a time window the last component in which user made a fixation in the previous time window.\\
        \\
        \hline
        \\
        $\lbrace [Component]\_end\_r1\rbrace_{c}$ & Similar to $\lbrace AOI\_end\_r1 \rbrace_{j}$ features, but per component, built from the feature $[Component]\_end$.\\
        \\
        \hline
        \\
        $\lbrace [Component]\_tslv \rbrace_{c}$ &The features ``time since last visit'' for each component is similar to $\lbrace AOI\_{tslv} \rbrace_{j}$.\\
        \\
        \hline
        \\
        $\lbrace [Component]\_atv \rbrace_{c}$ & Features ``average time of visits'' for each component give the average time in milliseconds of the fixations made on the component by user in his interaction with the web page. If there are no fixations in that component, it is assigned the value $0$.\\
        \\
        \hline
        \\
        $\lbrace [Component]\_atbv \rbrace_{c}$ & The ``average time between visits'' features for each component represent the average time in milliseconds that it has taken the user to perform a fixation on the component since last fixation on the same component. It is assigned the value $0$ in case there are no fixations. \\
        \\
        \hline
        \\
        $\lbrace Heat\_AOI \rbrace_{j}$ & Using the signals registered in the training set, a feature of frequency of visualizations is generated in each one of the defined AOI's, in the time windows made by other users $v_j^{(t)}$ in the times corresponding to the analyzed time window $v^{(t)}$, that is:\\
         & \multicolumn{1}{c}{$Heat\_AOI(v^{(t)}) = \frac{\sum_{u=1}^{U_{train}}{AOI(v_u^{(t)})}}{U_{train}}$}\\
         \\
        \hline
\end{tabular}
\label{variables}

\end{table*}

The feature ``Heat AOI'' adds information about the behavior of other users to the behavior of a particular participant. It is expected that for a stimulus, the model will learn to recognize the visual attention patterns of users over time, at the population level. In this way, the weighting of this feature in the classification models allows learning how to use these patterns in predictions.

Subsequently, \textbf{visual kinematics} features are defined, shown in table \ref{variablesCor}. Position, velocity and acceleration measurements are calculated in the X and Y coordinates of the gaze position over the stimulus. These features seek to incorporate information in relation to the shape of the ocular movement carried out in a period of time before the beginning of the window, which would allow knowing the future movement patterns. The definitions for the Y coordinate features ($coordY$, $MeanY$, $StdY$, $VelY$, $MeanVelY$, $StdVelY$, $AclY$, $MeanAclY$ and $StdAclY$) are similar to those of the X coordinate case.

\begin{table*}
    \centering
    \caption{Features of visual kinematics}
    \begin{tabular}{c|p{14 cm}}
    
    \hline
    \multicolumn{1}{c}{\textbf{Feature}} & \multicolumn{1}{c}{\textbf{Definition}}\\
    \hline
    \\
    $coordX$ & It is considered the last position of the previous time window, before starting the time window, registered for the user. \\
    \\
    \hline
    \\
    $MeanX$, $StdX$ & Mean and standard deviation of the gaze position in the X coordinate recorded in the user interaction with web page in the previous time window.\\
    \\
    \hline
    \\
    $VelX$ & Derivative from position data at X coordinate. This feature takes the velocity value corresponding to the last register of the previous time window. \\
    \\
    \hline
    \\
    $MeanVelX$ & \multirow{2}{=}{\setlength\parskip{\baselineskip}%
    Mean and standard deviation of the speed in X, of the data recorded in the previous time window.} \\
    $StdVelX$ & \\
    \\
    \hline
    \\
    $AclX$ & Derivative from speed data at X coordinate. This feature takes the value corresponding to the last acceleration register during navigation. \\
    \\
    \hline
    \\
    $MeanAclX$ & \multirow{2}{=}{\setlength\parskip{\baselineskip}%
    Mean and standard deviation of the acceleration data in the X coordinate recorded in the previous time window.} \\
    $StdAclX$ & \\
    \\
    \hline
\end{tabular}
\label{variablesCor}

\end{table*}

Finally, features related to the \textbf{eye-tracking function}, both fixation and saccade, are added (see table \ref{variablesOcu}). These features capture the areas to be visited according to the ocular movements registered for the user.

\begin{table*}
    \centering
    \caption{Features in Eye-Tracking Function}
    \begin{tabular}{c|p{14 cm}}
    
    \hline
    \multicolumn{1}{c}{\textbf{Feature}} & \multicolumn{1}{c}{\textbf{Definition}}\\
    \hline
    \\
    $NFix$ & Number of fixations made by the user during his interaction with web page along time.\\
    \\
    \hline
    \\
    $TPromFix$ & \multirow{3}{=}{\setlength\parskip{\baselineskip}%
    Average time, maximum time and minimum duration of the fixations made by the user during navigation.}\\
    $TMaxFix$ & \\ 
    $TMinFix$ & \\
    \\
    \hline
    \\
    $NSac$ & Total number of saccades made by the user up to the time before the start of the current time window.\\
    \\
    \hline
    \\
    $APromSac$ & \multirow{3}{=}{\setlength\parskip{\baselineskip}%
    Average, maximum and minimum amplitude of the saccades taken by the user up to the moment before the beginning of the current window.}\\
    $AMaxSac$ & \\
    $AMinSac$ & \\
    \hline
    
\end{tabular}
\label{variablesOcu}

\end{table*}

\subsection{Classification Models}

Five feature selection methods are executed, namely, maximization of mutual information (MLMIM), join mutual information (MLJMI) \cite{sechidis2014information}, min-redundancy max-relevance (MLMRMR) \cite{peng2005feature}, normalized mutual information feature selection (MIFS) \cite{jian2016multi} and robust feature selection (RFS) \cite{nie2010efficient}, on multilabel classification methods chosen because of their wide use in the literature. The configurations used are briefly described below:

\textbf{Ridge Regression (RR):} This model is used through the multilabel classification approaches of the binary relevance and classification chain, choosing the number of features in the range $5:5:40$, through the methods of feature selection mentioned. Given the features, different values are chosen for the parameter $\lambda = 0.25:0.25:2$, and then $\beta$ parameter is estimated.

\textbf{K-nearest neighbor (KNN):} This method is used with the binary relevance and classification chain approaches, setting the number of features to use in the range $5:5:40$ and the values of number of neighbors in the range $K=5:5:30$.

\textbf{Support vector machine (SVM):} This method is used with the binary relevance and classification chain approaches, setting the number of features to be used in the range $5:5:40$, chosen through the feature selection methods, and the parameter $C=0.2:0.2:2$.

\textbf{Multilabel K-nearest neighbor (MLKNN):} It is used directly without the approaches of binary relevance or the classification chain, setting the number of features to use in the range $5:5:40$, chosen through the methods of feature selection, and the parameter relative to the number of neighbors as $K=15$.

\subsection{Performance Metrics}

Models are evaluated by K-fold cross-validation, so the evaluation metrics are independently averaged between the k training and testing sets. 
To measure the quality of the solution and choose the best configuration, the following metrics are used. $Y_{i}$ corresponds to the real label vector of the visit intention components, and $Z_{i}$ corresponds to the prediction visit intention vector. Only in this case does $n$ represent the number of instances to evaluate.

The problem is originally seen as the search for a user's attention zones over time. This approach allows leeway such that the model does not have to achieve the full prediction of the area served by the user.
For this reason, the recall metric is not used directly for the evaluation of a model, which establishes the percentage of activation in the labels that are predicted, but it is used harmoniously in the F-measure metric. On the other hand, it is expected that in cases where the model establishes tag activation, these will actually be areas served by users, minimizing the error to the maximum amount. For this reason, the precision metric, which measures the percentage of activation occurrences that actually happen within the total activation proposed by the model, is used directly in the model selection.

\textbf{Area under the curve (AUC):} A receiver operating characteristic (ROC) curve demonstrates a classifier's ability to discriminate between positive and negative values by changing a threshold value. When the area below the curve is one, it represents a perfect prediction. In the multilabel classification case, it can be calculated from the macro approach, using the rankings of the instances for each label. Ranking $r(x_i, l)$ is defined as the function for which the instance $x_{i}$ and the label $y_{l}$ returns the degree of confidence of l in the prediction $Z_{i}$. In this way, the calculation is given by:

$$AUC = \frac{1}{k}\sum_{l \in \mathcal{L}}\frac{|x^{\prime},x^{\prime \prime}:r(x^{\prime},y_{l}) 	\geq r(x^{\prime \prime}, y_{l}),(x^{\prime},x^{\prime \prime}) \in A|}{|X_{l}||\bar{X}_{l}|}$$

$$X_{l}=\lbrace x_{i}|y_{l}\in Y_{i}\rbrace,\ \bar{X}_{l}=\lbrace x_{i}|y_{l} \not\in Y_{i} \rbrace, \ A=X_{l}\times \bar{X}_{l}$$

\textbf{Accuracy (Acc):} This is the ratio between the number of correctly predicted labels and the actually active labels, given individually by components. It is averaged across all instances:

$$Accuracy= \frac{1}{n}\sum_{i=1}^{n}\frac{|Y_{i} \cap Z_{i}|}{|Y_{i} \cup Z_{i}|}$$

\textbf{Subset Accuracy (Exact):} This is the instances percentage, where all coordinates of each vector that are correctly labeled are compared to the total number of labels (percentage of instances correctly predicted) \footnote{Operator $\llbracket\ \rrbracket$ corresponds to the bracket of Iverson (1 if the logical proposition is true; 0 if not)}.

$$Subset Accuracy = \frac{1}{n}\sum_{i=1}^{n} \llbracket Y_{i}=Z_{i} \rrbracket$$

\textbf{Precision (Pre):} The ratio between the number of correctly classified labels and the total number of labels. Intuitively, it corresponds to the percentage of labels predicted as true and that are really important for the instance.

$$Precision=\frac{1}{n}\sum_{i=1}^{n}\frac{|Y_{i}\cap Z_{i}|}{|Z_{i}|}$$

\textbf{F-measure (F-mea):} It is the average between precision and recall calculated in a harmonic way. It measures how many relevant tags are predicted and how many of the predicted labels are relevant.

$$ F-measure = 2\cdot \frac{Precision\cdot Recall}{Precision+Recall}$$

where Recall is the percentage of correctly predicted labels among all labels:

$$ Recall = \frac{1}{n} \sum_{i=1}^{n} \frac{|Y_{i} \cap Z_{i}|}{|Y_{i}|}$$

\section{Experimental Results} \label{ExperimentalResult}

Classification models are executed with different configurations, making a cross-validation of 10 folds. The average results measured in the test sets are given in \ref{resultados}. Test set samples are not balanced, since synthetic data classification is not relevant. The results therefore correspond to the original distributions of the labels.

\subsection{Principal Results} \label{resultados}

The proposed configurations are executed using the programming given in \cite{kimura2017mlc} (``MLC Toolbox''). The results are presented in table \ref{tab:resultados}, where the best models are summarized (choosing the number of features and parameters) as detailed below:

\begin{itemize}
    \item \textbf{Ridge Regression Binary Relevance (RR-BR):} Best results are achieved with the RFS feature selection method, using 30 predictive features and the parameter $\lambda=1.75$ (weighting of the coefficients of the variables in the target function).
    \item \textbf{Ridge Regression Classification Chain (RR-CC):} Best results are achieved with the RFS feature selection method, using 30 predictive features and the parameter $\lambda=1.75$ (weighting of the coefficients of the variables in the target function).
    \item \textbf{KNN Binary Relevance (KNN-BR):} Best results are achieved with the MLMIM feature selection method, using 10 predictive features and the parameter $K=20$ (neighborhood size).
    \item \textbf{KNN Classification Chain (KNN-CC):} Best results are achieved with the MLJMI feature selection method, using 10 predictive features and the parameter $K=15$ (neighborhood size).
    \item \textbf{SVM Binary Relevance (SVM-BR):} Best results are achieved with the MLMIM feature selection method, using 40 predictive features and the parameter $C=0.2$ (weighting of the loss function in the target function).
    \item \textbf{SVM Classification Chain (SVM-CC):} Best results are achieved with the RFS feature selection method, using 30 predictive features and the parameter $C=0.4$ (weighting of the loss function in the target function). 
    \item \textbf{MLKNN} Best results are achieved with the MLMIM feature selection method, using 10 predictive features.
\end{itemize}

\begin{table*}
    \centering
    \caption{Results of average metrics in the 10-folds are delivered. The standard deviation of these values within the same sets is given in parentheses.}
    \begin{tabular}{c|ccccc|cccccc}
    \hline
          & \multicolumn{5}{c|}{\textbf{Metrics}} & \multicolumn{6}{c}{\textbf{AOI prediction}} \\
    \hline
    \textbf{Model} & \textbf{AUC} & \textbf{Exact} & \textbf{Fscore} & \textbf{Acc} & \textbf{Pre} & \textbf{AOI 1} & \textbf{AOI 2} & \textbf{AOI 3} & \textbf{AOI 4} & \textbf{AOI 5} & \textbf{AOI 6} \\
    \hline
    \multirow{2}[2]{*}{\textbf{RR-BR}} & 0.844 & 0.285 & 0.658 & 0.566 & 0.877 & 0.755 & 0.686 & 0.747 & 0.725 & 0.785 & 0.971 \\
          & (0.006) & (0.010) & (0.011) & (0.010) & (0.007) & (0.016) & (0.012) & (0.009) & (0.012) & (0.013)  & (0.006) \\
    \hline
    \multirow{2}[2]{*}{\textbf{RR-CC}} & 0.844 & 0.285 & 0.658 & 0.556 & 0.877 & 0.755 & 0.686 & 0.747 & 0.725 & 0.785 & 0.971 \\
          & (0.006) & (0.010) & (0.011) & (0.010) & (0.007) & (0.016) & (0.012) & (0.009) & (0.012) & (0.013)  & (0.006) \\
    \hline
    \multirow{2}[2]{*}{\textbf{KNN-BR}} & \textbf{0.843} & \textbf{0.303} & \textbf{0.670} & \textbf{0.571} & \textbf{0.879} & \textbf{0.769} & \textbf{0.702} & \textbf{0.760} & \textbf{0.730} & \textbf{0.782} & \textbf{0.972} \\
          & \textbf{(0.005)} & \textbf{(0.008)} & \textbf{(0.009)} & \textbf{(0.008)} & \textbf{(0.004)} & \textbf{(0.019)} & \textbf{(0.01)} & \textbf{(0.008)} & \textbf{(0.01)} & \textbf{(0.013)}  & \textbf{(0.004)} \\
    \hline
    \multirow{2}[2]{*}{\textbf{KNN-CC}} & 0.839 & 0.302 & 0.679 & 0.579 & 0.878 & 0.763 & 0.710 & 0.729 & 0.790 & 0.79  & 0.967 \\
          & (0.005) & (0.006) & (0.007) & (0.006) & (0.003) & (0.015) & (0.008) & (0.01) & (0.01) & (0.013)  & (0.004) \\
    \hline
    \multirow{2}[2]{*}{\textbf{SVM-BR}} & \multirow{2}[2]{*}{-} & 0.291 & 0.663 & 0.564 & \multirow{2}[2]{*}{-} & 0.777 & 0.681 & 0.757 & 0.712 & 0.782 & 0.969 \\
          &       & (0.012) & (0.005) & (0.013) &       & (0.014) & (0.01) & (0.008) & (0.011) & (0.013)  & (0.008) \\
    \hline
    \multirow{2}[1]{*}{\textbf{SVM-CC}} & \multirow{2}[1]{*}{-} & 0.287 & 0.662 & 0.561 & \multirow{2}[1]{*}{-} & 0.775 & 0.682 & 0.751 & 0.712 & 0.775 & 0.970 \\
          &       & (0.009) & (0.009) & (0.009) &       & (0.015) & (0.009) & (0.007) & (0.011) & (0.014)  & (0.006) \\
    \hline
    \multirow{2}[1]{*}{\textbf{MLKNN}} & 0.844 & 0.301 & 0.675 & 0.574 & 0.882 & 0.767 & 0.708 & 0.756 & 0.732 & 0.781 & 0.97 \\
          & (0.005) & (0.009) & (0.009) & (0.009) & (0.005) & (0.017) & (0.01) & (0.007) & (0.011) & (0.013) & (0.004) \\
    \hline
    
    \end{tabular}
\label{tab:resultados}

\end{table*}

The best result is obtained with the model KNN binary relevance (model KKN-BR), with parameter $K=20$ (number of neighbors considered by the method) and the selection of 10 features by the MLMIM method, since this maximizes the accuracy and subset accuracy metrics. The features chosen are as follows: 

\begin{itemize}
    \item \textbf{Ocular fixation model feature}: $AOI\_6\_r1$, $AOI\_1\_end\_r1$, $AOI\_3\_end\_r1$, $AOI\_6\_end\_r1$, $AOI\_2\_end\_r1$,  $AOI\_3\_r1$, $AOI\_1\_r1$ AOI ocular fixations are selected.
    \item \textbf{Visual kinematics features}: $coordY$, $coordX$, $MeanY$
    \item \textbf{Eye-tracking function features}: No variables are selected.
\end{itemize}

\begin{figure}
\centering
\includegraphics[width=0.4\textwidth]{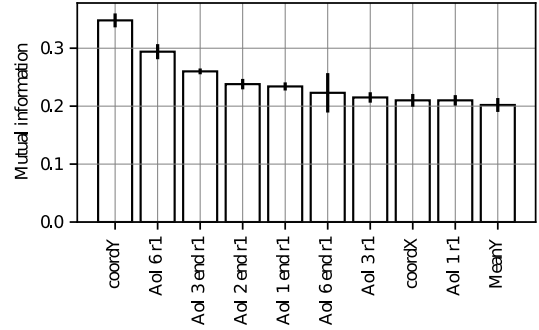}
\caption{Mutual information for each selected feature by the MLMIM method.} \label{MutualInformation}
\end{figure}

Figure \ref{MutualInformation} graphs the approximation made by the MLMIM method for the calculation of discretized (floor approximation of values) mutual information between $X$ (features) and $Y$ (labels), for variables selected across all folds. Higher values indicate a better relation of the variables with respect to the dependent variable. Possible values are in range $0 \leq I(X;Y) \leq min(H(X),H(Y))$ for each value, where $H()$ is the entropy and $I(X;Y)$ is mutual information value.


Ten features are listed. It is observed that all sets require the use of visual kinetics features of gaze position at the beginning of the time window. In addition, it is observed that the average Y coordinate recorded in the past is sometimes required. Subsequently, it is required to know the activation of any of the AOIs in the previous time window, or the information about the AOI in which the previous window ends.

It is interesting to consider that the model only requires knowing the activation of AOIs in a window of time in the past and not the full activity of user interaction. In the sets in which component information is required, the use of the features that indicate the activation of the dynamic components of the website in the previous time window is observed, that is, the menu bar and/or the pop-up registration.

Studying the behavior in each of the AOI's (right part of table \ref{tab:resultados}), it is observed that the model chosen presents the best accuracy values for three AOIs and approaches the best results of the other models for the other AOI's.

The results found are validated by comparing them with the biases towards the majority class, which is the proportion of labels with more repetitions with respect to the total. In the formula, $m$ is sample number, $MC$ is majority class and the function inside the argmax calculation is the majority class ratio.

$$MC \ (AOI_{j}) = \argmax_{x=\lbrace 0,1 \rbrace}\ \left( \frac{\sum_{q=1}^{m}\mathbbm{1}(AOI_{j}^{(q)}=x)}{m} \right)$$
\\
For each AOI, the $MC$ ratio values are 0.653, 0.567, 0.513, 0.629, 0.727, and 0.954, for AOI's $1,\dots,6$, where the majority class corresponds to the activation of each AOI (1 label), except in the case of AOIs 2 and 3 (0 label). At the same time, the bias to the majority class when considering the complete instance is 0.133 (considering the combination of labels of each AOI of a vector as a label), which corresponds to the activation of AOIs 2 and 3. Therefore, at the individual level and as a whole, the model improves the trivial case of always choosing the majority class.

\subsection{Sensitivity Analysis by Time Window Lengths} \label{sensibilizacion}

Sensitivity analysis is carried out, given variations in the time windows' duration to be considered in classification. With this analysis, it is expected that the robustness of the methodology can be evaluated and used for the prediction of visit intentions.

First, models are run with all the combinations again, considering the duration of time windows with $\tau = 3, 10, 15, 20$ seconds. Table \ref{tab:Sen_Models} shows the best settings for each prediction model. Table \ref{tab:Sen_Metrics} shows the metrics obtained by the best models for each combination. 

\begin{table}
    \centering
    \caption{Better configurations for each classification model according to time window duration. FS: Feature selection method, NF: Number of features, P: Parameters.}
    \begin{tabular}{cccccc}
    
    \hline
        & & \multicolumn{4}{c}{\textbf{Time Windows Duration}} \\
        \multicolumn{2}{c}{\textbf{Model}} & \textbf{$\tau = 3$} & \textbf{$\tau = 10$} & \textbf{$\tau = 15$} & \textbf{$\tau = 20$} \\
    \hline
    \multirow{5}[1]{*}{\textbf{\shortstack[c]{RR\\BR}}} & \textbf{FS} & RFS   & MIFS  & RFS   & F-Score \\
          \cmidrule{2-6}
          & \textbf{NF} & 35    & 30    & 5     & 5 \\
          \cmidrule{2-6}
          & \textbf{P} & $\lambda=0.25$ & $\lambda=1$ & $\lambda=1$ & $\lambda=1.5$ \\
    \hline
    \multirow{5}[1]{*}{\textbf{\shortstack[c]{RR\\CC}}} & \textbf{FS} & RFS   & MLJMI & RFS   & F-Score \\
        \cmidrule{2-6}
          & \textbf{NF} & 35    & 40    & 5     & 5 \\
        \cmidrule{2-6}
          & \textbf{P} & $\lambda=0.25$ & $\lambda=0.25$ & $\lambda=1$ & $\lambda=1.5$ \\
    \hline
    \multirow{5}[1]{*}{\textbf{\shortstack[c]{KNN\\BR}}} & \textbf{FS} & MLMRMR & MLMIM & F-Score & F-Score \\
        \cmidrule{2-6}
          & \textbf{NF} & 10    & 40    & 25    & 10 \\
        \cmidrule{2-6}
          & \textbf{P} & $K=20$  & $K=30$  & $K=25$  & $K=30$ \\
    \hline
    \multirow{5}[1]{*}{\textbf{\shortstack[c]{KNN\\CC}}} & \textbf{FS} & MLJMI & MLJMI & RFS   & F-Score \\
        \cmidrule{2-6}
          & \textbf{NF} & 25    & 35    & 20    & 10 \\
        \cmidrule{2-6}
          & \textbf{P} & $K=10$  & $K=10$  & $K=25$  & $K=25$ \\
    \hline
    \multirow{5}[1]{*}{\textbf{\shortstack[c]{SVM\\BR}}} & \textbf{FS} & MLMIM & F-Score & F-Score & F-Score \\
        \cmidrule{2-6}
          & \textbf{NF} & 40    & 40    & 5     & 10 \\
        \cmidrule{2-6}
          & \textbf{P} & $C=0.2$ & $C=0.2$ & $C=0.4$ & $C=0.2$ \\
    \hline
    \multirow{5}[1]{*}{\textbf{\shortstack[c]{SVM\\CC}}} & \textbf{FS} & RFS   & MIFS  & RFS   & F-Score \\
        \cmidrule{2-6}
          & \textbf{NF} & 35    & 40    & 5     & 5 \\
        \cmidrule{2-6}
          & \textbf{P} & $C=1.6$ & $C=0.4$ & $C=2$ & $C=0.4$ \\
    \hline
    \multirow{5}[1]{*}{\textbf{\shortstack[c]{ML-\\KNN}}} & \textbf{FS} & MLJMI & MLJMI & RFS   & RFS \\
        \cmidrule{2-6}
          & \textbf{NF} & 15    & 40    & 20    & 30 \\
        \cmidrule{2-6}
          & \textbf{P} & $K=15$  & $K=15$  & $K=15$  & $K=15$ \\
    \hline
\end{tabular}
\label{tab:Sen_Models}

\end{table}

\begin{table}
    \centering
    \caption{Sensitivity analysis metrics on the duration of time windows for the best models.}
    \begin{tabular}{cc|ccccc}
    
    \hline
    & & \multicolumn{5}{c}{\textbf{Time Windows Duration (sec.)}} \\
    \multicolumn{2}{c}{\textbf{Model}} & \textbf{$\tau=3$} & \textbf{$\tau=5$} & \textbf{$\tau=10$} & \textbf{$\tau=15$} & \textbf{$\tau=20$} \\
    \hline
    
    \multirow{5}[0]{*}{\textbf{\shortstack[c]{RR\\BR}}} & \textbf{AUC} & 0.876 & 0.844 & 0.820 & \textbf{0.788} & \textbf{0.792} \\
          & \textbf{Exact} & 0.416 & 0.285 & 0.198 & \textbf{0.271} & \textbf{0.363} \\
          & \textbf{Fscore} & 0.683 & 0.658 & 0.719 & \textbf{0.781} & \textbf{0.850} \\
          & \textbf{Acc} & 0.609 & 0.566 & 0.606 & \textbf{0.686} & \textbf{0.771} \\
          & \textbf{Pre} & 0.893 & 0.877 & 0.881 & \textbf{0.861} & \textbf{0.913} \\
    \hline
    \multirow{5}[1]{*}{\textbf{\shortstack[c]{RR\\CC}}} & \textbf{AUC} & 0.876 & 0.844 & 0.825 & 0.788 & 0.792 \\
          & \textbf{Exact} & 0.416 & 0.285 & 0.192 & 0.271 & 0.363 \\
          & \textbf{Fscore} & 0.683 & 0.658 & 0.728 & 0.781 & 0.850 \\
          & \textbf{Acc} & 0.609 & 0.556 & 0.614 & 0.686 & 0.771 \\
          & \textbf{Pre} & 0.893 & 0.877 & 0.885 & 0.861 & 0.913 \\
    \hline
    \multirow{5}[2]{*}{\textbf{\shortstack[c]{KNN\\BR}}} & \textbf{AUC} & 0.865 & \textbf{0.843} & 0.824 & 0.813 & 0.804 \\
          & \textbf{Exact} & 0.423 & \textbf{0.303} & 0.194 & 0.233 & 0.343 \\
          & \textbf{Fscore} & 0.687 & \textbf{0.670} & 0.733 & 0.789 & 0.845 \\
          & \textbf{Acc} & 0.613 & \textbf{0.571} & 0.621 & 0.687 & 0.763 \\
          & \textbf{Pre} & 0.889 & \textbf{0.879} & 0.887 & 0.895 & 0.908 \\
    \hline
    \multirow{5}[2]{*}{\textbf{\shortstack[c]{KNN\\CC}}} & \textbf{AUC} & \textbf{0.865} & 0.839 & \textbf{0.815} & 0.801 & 0.800 \\
          & \textbf{Exact} & \textbf{0.424} & 0.302 & \textbf{0.200} & 0.246 & 0.360 \\
          & \textbf{Fscore} & \textbf{0.696} & 0.679 & \textbf{0.711} & 0.775 & 0.844 \\
          & \textbf{Acc} & \textbf{0.621} & 0.579 & \textbf{0.601} & 0.674 & 0.764 \\
          & \textbf{Pre} & \textbf{0.888} & 0.878 & \textbf{0.881} & 0.880 & 0.909 \\
    \hline
    \multirow{5}[2]{*}{\textbf{\shortstack[c]{SVM\\BR}}} & \textbf{AUC} & -     & -     & -     & -     & - \\
          & \textbf{Exact} & 0.402 & 0.291 & 0.174 & 0.176 & 0.169 \\
          & \textbf{Fscore} & 0.686 & 0.663 & 0.669 & 0.722 & 0.701 \\
          & \textbf{Acc} & 0.609 & 0.564 & 0.554 & 0.612 & 0.601 \\
          & \textbf{Pre} & -     & -     & -     & -     & - \\
    \hline
    \multirow{5}[1]{*}{\textbf{\shortstack[c]{SVM\\CC}}} & \textbf{AUC} & -     & -     & -     & -     & - \\
          & \textbf{Exact} & 0.412 & 0.287 & 0.200 & 0.263 & 0.305 \\
          & \textbf{Fscore} & 0.690 & 0.662 & 0.680 & 0.758 & 0.787 \\
          & \textbf{Acc} & 0.613 & 0.561 & 0.568 & 0.660 & 0.702 \\
          & \textbf{Pre} & -     & -     & -     & -     & - \\
    \hline
    \multirow{5}[1]{*}{\textbf{\shortstack[c]{ML-\\KNN}}} & \textbf{AUC} & 0.871 & 0.844 & 0.823 & 0.805 & 0.809 \\
          & \textbf{Exact} & 0.420 & 0.301 & 0.193 & 0.244 & 0.365 \\
          & \textbf{Fscore} & 0.694 & 0.675 & 0.731 & 0.782 & 0.839 \\
          & \textbf{Acc} & 0.618 & 0.574 & 0.619 & 0.681 & 0.759 \\
          & \textbf{Pre} & 0.891 & 0.882 & 0.885 & 0.883 & 0.903 \\
    \hline

    \end{tabular}
    \label{tab:Sen_Metrics}
\end{table}

It can be seen that the settings chosen for each model are not the same for all the values of the temporary duration of the window ($\tau$). The best results are obtained with the use of an RR-BR model for longer time windows. It is interesting to see that the case of $\tau = 3$ obtains a better Exact metric.

\begin{figure} [H]
\centering
\includegraphics[]{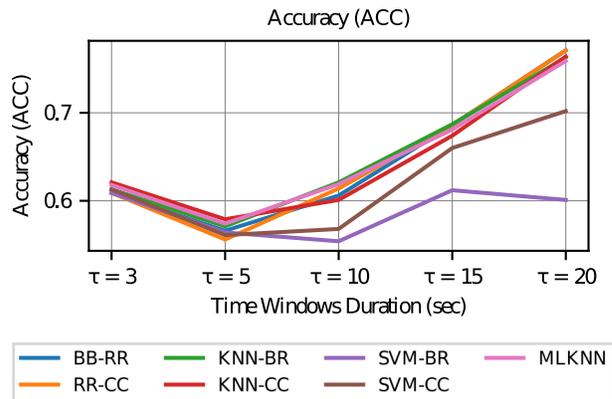}
\caption{Accuracy vs. Time Window Duration.} \label{sen_graficoACC}
\end{figure}

\begin{figure} [H]
\centering
\includegraphics[]{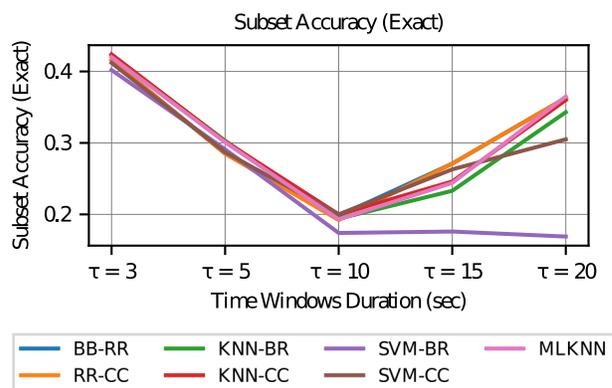}
\caption{Subset Accuracy vs. Time Window Duration.} \label{sen_graficoSACC}
\end{figure}

First, the values of the subset accuracy (Exact) and accuracy (Acc) metrics are inspected for different values of $\tau$. Figure \ref{sen_graficoACC} and \ref{sen_graficoSACC} shows the graphs of these values for each classification model configuration. 

From these metrics, it can be seen that if a model with the best estimates of the activation of each AOI is desired, then the case of a longer duration in the time windows ($\tau = 20$) is preferred, since in this case the accuracy (which measures performance by the individual label of each AOI) increases in all models. However, when studying the subset accuracy, that is, the correctness in the estimation of the activation of labels in a vector way (all the AOI's within a time window), the duration of the most extreme time windows is preferred, that is, this metric increases with time windows of shorter duration ($\tau = 3$) or longer duration ($\tau = 20$). 

To analyze the above, it can be considered that if all models achieve the same level of training (which is not necessarily true), the vectors of the visit intention in smaller and larger time windows show less variability. As the size of the window increases to infinity, the variability in the visit intention tends to zero since it is expected to take the value 1 in all AOI's (although this may not be true in cases where AOI 6 is not visited). In the case of a very small window (tending to zero), the intention to visit all AOI's should tend to zero as there is no time to make transitions to other AOI's.

\begin{figure} [H]
\centering
\includegraphics[]{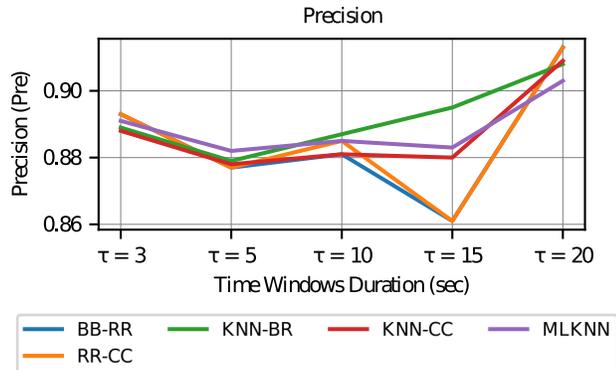}
\caption{Precision vs. Time Window Duration.} \label{sen_grafico2}
\end{figure}

Subsequently, to better understand the results, the precision is studied. In figure \ref{sen_grafico2}, it is observed that the values for precision appear to be stable according to the lengths of the windows, with the exception of the classifiers based on ridge regression, which tend to decrease slightly more when choosing $\tau = 15$.

In all cases, precision values greater than 0.8 are achieved, a value commonly considered to be good in the literature. This result means that in the label activation predicted by the model, more than 80\% of the cases are performed correctly. In the case of $\tau = 20$, it achieves better accuracy, which may be related to the fact that there are more active labels (as in the case of windows tending to infinity discussed in the previous paragraph), which causes the existence of a greater number of true positive values.

\begin{figure} [H]
\centering
\includegraphics[width=0.45\textwidth]{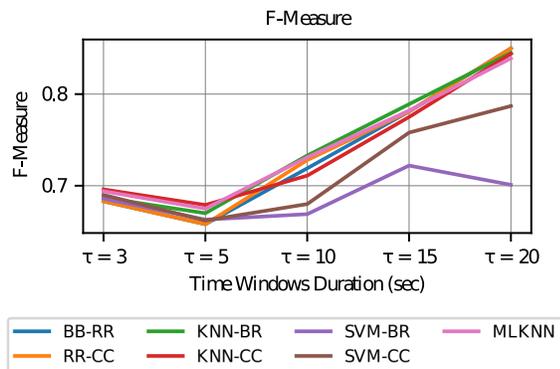}
\caption{F-Measure vs. Time Window Duration.} \label{sen_grafico3}
\end{figure}

In the case of the F-measure values (see figure \ref{sen_grafico3}), a high trend is observed as the duration of the time windows increases. This result indicates that for longer time windows, greater harmony is achieved between precision and recall.

\begin{figure} [H]
\centering
\includegraphics[]{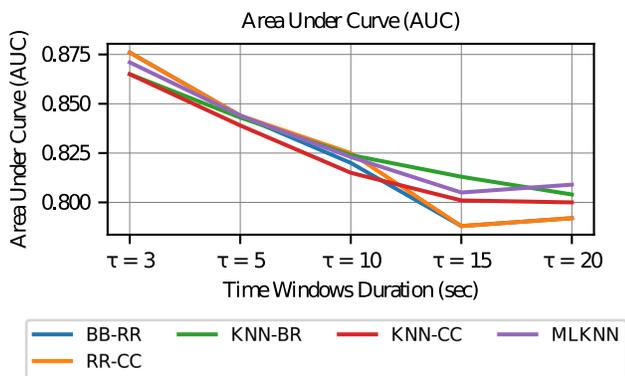}
\caption{Area Under Curve vs. Time Window Duration.} \label{sen_grafico4}
\end{figure}

Finally, when evaluating the area under the curve (AUC) graph in figure \ref{sen_grafico4}, it can be observed that it decays for values with longer time windows.
This observation indicates that the number of times that the ranking value (confidence given to a certain label) of the positive labels (corresponding to the activation of an AOI) exceeds the ranking for a negative label (that is, no activation of an AOI) is lower. Therefore, the power of discrimination between values predicted as positive (true positives) and negative values that are classified as positive (false positive) is lower in longer time windows.

The drop in AUC values shows that longer time window models have worse discrimination between classes. However, in all cases, values above 0.8 are obtained, which is considered good in the literature.

Given the above, it is concluded that at longer time windows, the accuracy and subset accuracy will tend to improve, which does not necessarily reflect a good model performance. It is necessary to consider the discriminative power between classes of the model, for example, from metrics such as AUC. On the other hand, it is important to consider the exploratory analysis of the labels (individual and vectorial) in each case ($tau$ chosen), since it is necessary to make a trade-off between the variability of the labels and the time used to make the predictions.

From previous results, it is considered that the estimation of shorter windows is better, since it gives more detailed information of the visual attention in a better discretized time. However, to ensure that the results of the model are valuable, the use of $\tau = 3$ must be evaluated both with respect to the time required by the model for its execution (which does not exceed the duration of the corresponding window) and the variability of the label sufficiency. 
By the previous criteria, the model from the beginning was considered with $\tau = 5$ as the main model.

\subsection{Time Overhead Calculation} \label{TimeOverhead}

The window processing time for the multilabel classification model chosen is given by the signal processing used in the prediction (cleaning of the signals and characteristics calculation), in addition to the execution of the classification.

It is interesting to study the classification times given the duration of the time windows, since on the same equipment, the methodology of the algorithm indicates the speed of computation. The processing times of characteristics will depend on the amount chosen by the model in each of the classifiers considered, according to the duration of the windows, in addition to the processing equipment.

To compare classifiers' performance, all combinations are run to classify windows of different time duration. In particular, windows of length $\tau = \lbrace 3, 10, 15, 20 \rbrace$ seconds are considered, in addition to the case of time windows of 5 seconds. Table \ref{tab:TimeOverhead} shows the mean and standard deviation of execution times of the classifiers in seconds. 

\begin{table*}
    \centering
    \caption{Time processing of the classification models for different window lengths.}
    \begin{tabular}{c|c|cc|cc|cc|cc|cc}
    
    \hline
    \multicolumn{1}{r}{} & \multicolumn{1}{r}{} & \multicolumn{2}{c}{\textbf{Tau = 3}} & \multicolumn{2}{c}{\textbf{Tau = 5}} & \multicolumn{2}{c}{\textbf{Tau = 10}} & \multicolumn{2}{c}{\textbf{Tau = 15}} & \multicolumn{2}{c}{\textbf{Tau = 20}} \\
    \hline
    \multicolumn{1}{c}{\textbf{Model}} &       & \textbf{Mean} & \textbf{SD} & \textbf{Mean} & \textbf{SD} & \textbf{Mean} & \textbf{SD} & \textbf{Mean} & \textbf{SD} & \textbf{Mean} & \textbf{SD} \\
    \hline
    \multirow{2}[2]{*}{\textbf{RR-BR}} & \textbf{Train} & 3.008 & 0.085 & 1.380 & 0.051 & 3.694 & 0.145 & 1.497 & 0.060 & 0.402 & 0.004 \\
          & \textbf{Test} & 0.000 & 0.000 & 0.000 & 0.000 & 0.000 & 0.000 & 0.000 & 0.000 & 0.000 & 0.000 \\
    \hline
    \multirow{2}[2]{*}{\textbf{RR-CC}} & \textbf{Train} & 3.020 & 0.058 & 1.392 & 0.051 & 0.097 & 0.010 & 1.464 & 0.043 & 0.378 & 0.008 \\
          & \textbf{Test} & 0.000 & 0.000 & 0.000 & 0.000 & 0.000 & 0.000 & 0.000 & 0.000 & 0.000 & 0.000 \\
    \hline
    \multirow{2}[2]{*}{\textbf{KNN-BR}} & \textbf{Train} & 3.252 & 0.037 & 0.548 & 0.020 & 0.400 & 0.016 & 0.356 & 0.004 & 0.355 & 0.006 \\
          & \textbf{Test} & 1.294 & 0.007 & 0.481 & 0.012 & 0.092 & 0.004 & 0.027 & 0.002 & 0.016 & 0.002 \\
    \hline
    \multirow{2}[2]{*}{\textbf{KNN-CC}} & \textbf{Train} & 0.002 & 0.002 & 0.000 & 0.000 & 0.000 & 0.000 & 1.427 & 0.033 & 0.395 & 0.009 \\
          & \textbf{Test} & 1.302 & 0.007 & 0.436 & 0.004 & 0.123 & 0.004 & 0.111 & 0.014 & 0.020 & 0.002 \\
    \hline
    \multirow{2}[2]{*}{\textbf{SVM-BR}} & \textbf{Train} & 3.730 & 0.052 & 1.773 & 0.005 & 1.205 & 0.012 & 0.636 & 0.010 & 0.445 & 0.006 \\
          & \textbf{Test} & 0.013 & 0.007 & 0.008 & 0.003 & 0.000 & 0.000 & 0.002 & 0.002 & 0.002 & 0.002 \\
    \hline
    \multirow{2}[2]{*}{\textbf{SVM-CC}} & \textbf{Train} & 4.511 & 0.054 & 1.870 & 0.031 & 3.905 & 0.049 & 1.559 & 0.047 & 0.488 & 0.016 \\
          & \textbf{Test} & 0.002 & 0.002 & 0.005 & 0.002 & 0.003 & 0.002 & 0.000 & 0.000 & 0.002 & 0.002 \\
    \hline
    \multirow{2}[2]{*}{\textbf{MLKNN}} & \textbf{Train} & 0.000 & 0.000 & 0.538 & 0.008 & 0.000 & 0.000 & 1.442 & 0.042 & 1.992 & 0.083 \\
          & \textbf{Test} & 0.300 & 0.018 & 0.081 & 0.002 & 0.008 & 0.003 & 0.020 & 0.010 & 0.013 & 0.008 \\
    \hline
    
    \end{tabular}
    \label{tab:TimeOverhead}
\end{table*}

Execution times for ridge regression models are less than one millisecond in the test sets. This outcome is because the model is based on the storage of the weighting parameters of each variable and only requires a multiplication with the features of the test sets to obtain the result.

Most expensive computational models correspond to those using the KNN base classifier. This is because for a new window in the test set, its classification requires searching for the neighborhood in the training set according to the size given by the parameters of the model.

\begin{figure}
\centering
\includegraphics[width=0.45\textwidth]{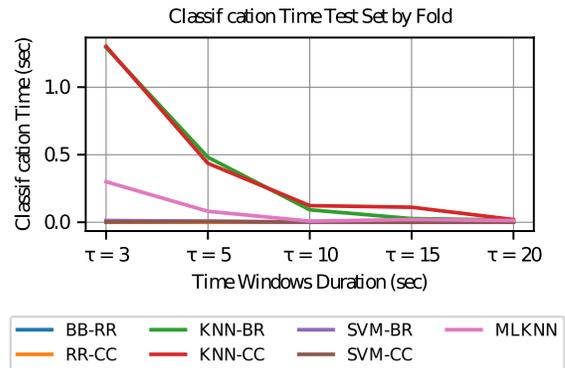}
\caption{Classification Time Test Set by Fold.} \label{graf_TimeOverhead}
\end{figure}

In the figure \ref{graf_TimeOverhead}, it is observed that the execution time decreases with increasing time window length; this effect occurs because with longer times, test sets are generated with a smaller number of temporary window samples.

Finally, it is observed that the classification times are less than the duration of each time window present in the test sets, so they are applicable in an online environment. This approach considers many windows, where the averages in the test sets are 398, 239, 120, 80, and 60 for duration $\tau$ of 3, 5, 10, 15 and 20 seconds, respectively.

\section{Discussion} \label{Discussion}
Thus far, a visual attention indicator called visit intention has been generated, with which it is possible to predict certain time intervals. The results obtained have been explained above. In this section, we will discuss some important points that should be considered about the development of this work.

The work has been developed in a ``controlled dynamic'' environment, that is, a web page in which the characteristics of structure and content are known, but that varies, under certain limits, according to the interaction that the user has with the page. In this way, a method of predicting the visit intention has been established without needing to know the structure of the site at all times. However, the fundamental idea is that the model learns to recognize the structure ``implicitly'' according to the data of the user's behavior and the compilation of the real values of visual attention in previous time windows.

The control of the website used is based on the structure of the elements (position within the screen of the static components) and functionality (position of the dynamic elements and possible changes of the site given by the user's clicks during their interaction with effects on the stimulus). At the same time, the experimental environment has been controlled, particularly in the lighting of the room in which the interaction takes place and the restricted physical movement of the users (through instruction of the experiment).

Given the experimental control, it is considered that this approach is not useful in predictions of visual attention carried out in egocentric vision environments discussed in Section \ref{egocentric}, since in these environments, the stimulus tends to be less controlled. In these environments, the temporal discretion of the stimulus, corresponding to the frames of a video, can suffer considerable variability, since what the user detects depends on (in addition to being given by the elements around him) his or her position, height and closeness (reference system from the observer).

A second aspect to consider is that the study is carried out for the case of bottom-up attention from the instructions of free navigation on a stimulus unknown by the user. However, within real web applications, it can be interesting to predict in top-down cases, that is, where the visual attention of a user is influenced by the task performed. The reason is because users of a site tend to look for certain information in specific areas, depending on their objectives. At the same time, their navigation in web environments depends on the knowledge about web design that each user possesses (learned based on interactions with other similar websites). In this way, the tasks present complications regarding the generalization of models. Depending on each task developed, models can be generated that consider this information as variables to generate new predictions.

On the other hand, the participant sample used in the experimentation corresponds to a convenience sample, in which users correspond to engineering students, for the most part, who would be expected to present a better performance in the use of technologies and to have a background regarding the design of the web pages. The study of results for other types of users would allow a better generalization of the models and, at the same time, could provide useful information regarding their improvement, for example, when considering clusters of similar users in the estimation of parameters. Moreover, this sample corresponds to people from the same location with shared cultural characteristics and native speech corresponding to that of the site studied.

Bearing in mind the above limitations, and although a specific website has been used, it is important to consider that these results have been obtained using a replicable methodology and with the use of calculable variables for any dynamic web stimulus and other visual stimuli in which its components could be identified. It is necessary to evaluate these models for other stimuli, which would ensure their generalization, depending on the quality of the results obtained. 

Regarding the scalability of the solution, it has been demonstrated that the execution of the classification models takes less time than the duration of the temporary windows that are being predicted. Additionally, the classification of multiple samples at the same time is considered (total of samples present in each test set). 
The transmission of data (architectures) and the calculation of features from the signals is a factor that must be considered in the selection of the temporary windows; however, in this study, these factors have not been studied in greater depth. 

Depending on the results in terms of the scalability study, the use of these approaches in the prediction would allow considering other types of platforms, such as applications in mobile web environments and tablets. This requires experimentation with new visual stimuli in a controlled manner in these new environments.

The multilabel classification models used allow the prediction of visit intention values (classification of the activation of each AOI in time windows corresponding to future times). In particular, the use of these models incorporates the information jointly to all the AOI's present in the stimulus, within each time window. In this way, the classification approaches used seem to be adequate in terms of understanding the visual attention of a user as a whole in all areas of the website. On the other hand, other multilabel classification models can be used, as can neural network models based on deep learning.

In this study, a subset of metrics has been chosen for the evaluation of each model and the comparison to select the best configurations. Both the accuracy and the subset accuracy give us information on the percentage of success of the model to predict values for each AOI and each set of active AOI's, respectively. Although in principle, these values are intended to be as high as possible, other metrics are considered to better understand the effects of the model. 

The precision metric is chosen given the approach adopted for the problem from experimentation, which seeks that the activations predicted by a model are as accurate as possible (increase the number of true positives), i.e., to be able to know which areas will be visited by a user with greater reliability. On the other hand, the recall metric frequently used in the literature has not been directly considered for model evaluation. However, it is used indirectly through the F-measure metric to check the ratio between the uncertainty in the predicted activations and the amounts of actual activations detected.
 Finally, it has been established that the area under the curve (AUC) allows checking the discriminative power between classes of each model. 

With respect to the main model, a multilabel classification approach has been used so that the prediction considers the information that relates them. However, when evaluating the results in each AOI separately, the accuracies correspond to 0.879, 0.769, 0.702, 0.760, 0.730, and 0.782 (see \ref{tab:resultados}), which yields an average \textbf{individual AOI accuracy of 0.77} for the model (not to be confused with ACC, which corresponds to the percentage of predictions of real activation without considering the prediction of the nonactivation of AOI). 

The results deliver a total of 10 characteristics in each of the training-test partitions (see figure \ref{MutualInformation}). It can be observed that in this model, no Eye Tracker feature function variables are used, and of the rest, 73\% correspond to eye fixation characteristics and 27\% to kinematic characteristics. 

The foregoing suggests that most of the prediction is made using the knowledge of the past behavior of the user, with both the transitions in the gaze and the areas of interest served. In particular, position kinematic variables are used, that is, the location of the user's gaze at the end of the previous time window. With this, the model is encouraged to learn the closest areas of interest for the user in a new time window based on the current position, which is useful for predicting the static AOI activations.
The variable ``coordY'' presents the greatest mutual information with the dependent variables, which is related to the fact that along the vertical axis, a greater variety is presented in the defined AOI in relation to the horizontal axis (variable ``coordX'' presents less mutual information with the dependent variables). 

In addition to knowing the starting point of a user's gaze, variables are used to record areas of interest in the previous window. It can be seen that the model requires knowing if the dynamic AOI's (AOI 1 and AOI 6) have been visited in that period. The second variable with more mutual information regarding the dependent variables corresponds to ``AOI\_6\_r1''; this shows that knowledge of the activation of the registration zone in the previous window allows the model to obtain information about the structure that can maintain the page in the time window to predict; and information is provided about whether the user will present new fixations in that zone. 
It should be noted that the model uses the activation information in the past (here, it requires knowledge of the stimulus) but does not need to generate predictions about the future structure or the time in which it will be maintained. The same applies in the case of ``AOI\_1\_r1'', but with a lower value of the mutual information presented with respect to the dependent variables.

Use is made of the variables ``AOI\_6\_end\_r1'' and ``AOI\_1\_end\_r1'', which correspond to knowing if the previous time window ends with the use of these dynamic AOI's, which unlike the variables ``AOI\_6\_r1'' and ``AOI\_1\_r1'' present term structure information from the previous time window directly (it is known that the user's gaze is on those AOI's at the beginning of the prediction window). It is expected that in most cases where a time window ends with the activation of a dynamic AOI, the activation will appear in the next window. 

Finally, the activation variables of AOI's 2 and 3 corresponding to the news sets present on the website are used. It is intuitively considered that these variables deliver spatial information to the model about the user's attention in a time window prior to which the prediction is sought to be made. Variables 3 and 4 are apparently are not necessary for the model.

Therefore, the proposed model complies with what is sought, i.e., through real-time recording of user behavior, predictions of areas of interest for future periods of time are achieved without the need to predict the structure and duration of these changes.

\section{Conclusion and Future Work} \label{Conclusions}

A predictive model is proposed for the concept of ``visit intention'', which provides information on the areas of interest of a website that will be visited by a user over a time period. It is not required to know the structure of the website in each time window, as in other studies found in the literature; rather, learning the dynamic changes through the gaze data registered for a user, estimating models through population behavior as characteristics and employing a user's visual kinetics to train individual models for each instance of prediction are utilized. The model is applicable from an analysis of execution times, and criteria have been given in the selection of the window duration used in the segmentation of the total time of user interaction, preferring for the use of windows of $\tau = 5$ seconds for this case.

As future work, it is expected that these results can be integrated in more advanced models of visual attention, in addition to the study of improvements based on models for segmentation of users and the use of deep learning methods. Furthermore, it is interesting to consider the study of the methodology of this style in other visual environments (e.g., mobile devices) and in cases related to specific tasks (e.g., search for a news item on the site).

\ifCLASSOPTIONcompsoc
  \section*{Acknowledgments}
\else
  \section*{Acknowledgment}
\fi

This work was partially financed by Conicyt Fondecyt 1191104, and Conicyt Fondecyt 1181809. The authors recognize the continuous support of ``Instituto Sistemas Complejos de Ingenier\'ia'' (Conicyt: Proyecto Basal FBO16). This article is part of the work of the e-mental health network of the University of Chile, VID U-REDES-C\_2018\_07.

\ifCLASSOPTIONcaptionsoff
  \newpage
\fi


\begin{thebibliography}{1}

\bibitem{bazzani2016recurrent}
Bazzani, L., Larochelle, H., Torresani, L.: Recurrent mixture density network for spatiotemporal visual attention. arXiv preprint arXiv:1603.08199. (2016)

\bibitem{guo2008spatio}
Guo, C., Ma, Q., Zhang, L.: Spatio-temporal saliency detection using phase spectrum of quaternion fourier transform. (2008)

\bibitem{zhao2015fixation}
Zhao, J., Siagian, C., Itti, L.: Fixation bank: Learning to reweight fixation candidates. In Proceedings of the IEEE Conference on Computer Vision and Pattern Recognition (pp. 3174-3182) (2015)

\bibitem{zhai2006visual}
Zhai Y., Shah M.: Visual attention detection in video sequences using spatiotemporal cues. In Proceedings of the 14th ACM international conference on Multimedia (pp. 815-824). ACM.(2006)

\bibitem{fang2014video}
Fang Y., Wang Z., Lin W., Fang Z.: Video saliency incorporating spatiotemporal cues and uncertainty weighting. IEEE Transactions on Image Processing, 23(9), 3910-3921. (2014)

\bibitem{sitzmann2018saliency}
Sitzmann V., Serrano A., Pavel A., Agrawala M., Gutierrez D., Masia B., Wetzstein, G.: Saliency in VR: How do people explore virtual environments?. IEEE transactions on visualization and computer graphics, 24(4), 1633-1642. (2018)

\bibitem{marmitt2002modeling}
Marmitt G., Duchowski A. T.: Modeling visual attention in VR: Measuring the accuracy of predicted scanpaths (Doctoral dissertation, Clemson University).(2002)


\bibitem{huang2018predicting}
Huang Y., Cai M., Li Z., Sato Y.: Predicting gaze in egocentric video by learning task-dependent attention transition. In Proceedings of the European Conference on Computer Vision (ECCV) (pp. 754-769). (2018)

\bibitem{zhang2017deep}
Zhang M., Teck Ma K., Hwee Lim J., Zhao Q., Feng, J.: Deep future gaze: Gaze anticipation on egocentric videos using adversarial networks. In Proceedings of the IEEE Conference on Computer Vision and Pattern Recognition (pp. 4372-4381). (2017)

\bibitem{shen2014webpage}
Shen C., Zhao Q.: Webpage saliency. In European conference on computer vision (pp. 33-46). Springer, Cham. (2014)

\bibitem{shen2015predicting}
Shen C., Huang X., Zhao Q.: Predicting eye fixations on webpage with an ensemble of early features and high-level representations from deep network. IEEE Transactions on Multimedia, 17(11), 2084-2093 (2015)

\bibitem{li2018webpage}
Li Y., Zhang Y.: Webpage Saliency Prediction with Two-stage Generative Adversarial Networks. arXiv preprint arXiv:1805.11374.(2018)

\bibitem{li2016webpage}
Li J., Su L., Wu B., Pang J., Wang C., Wu Z., Huang Q.: Webpage saliency prediction with multi-features fusion. In Image Processing (ICIP), 2016 IEEE International Conference on (pp. 674-678). IEEE. (2016)

\bibitem{xu2016spatio}
Xu P., Sugano Y., Bulling A.: Spatio-temporal modeling and prediction of visual attention in graphical user interfaces. In Proceedings of the 2016 CHI Conference on Human Factors in Computing Systems (pp. 3299-3310). ACM. (2016)

\bibitem{katsuki2014bottom}
Katsuki, F., Constantinidis, C.: Bottom-up and top-down attention: different processes and overlapping neural systems. The Neuroscientist, 20(5), 509-521. (2014)

\bibitem{rutishauser2004bottom}
Rutishauser, U., Walther, D., Koch, C., Perona, P.: Is bottom-up attention useful for object recognition?. In Proceedings of the 2004 IEEE Computer Society Conference on Computer Vision and Pattern Recognition, 2004. CVPR 2004. (Vol. 2, pp. II-II). IEEE. (2004, June)

\bibitem{connor2004visual}
Connor, C. E., Egeth, H. E., Yantis, S.: Visual attention: bottom-up versus top-down. Current biology, 14(19), R850-R852. (2004)

\bibitem{kruthiventi2016saliency}
Kruthiventi, S. S., Gudisa, V., Dholakiya, J. H., Venkatesh Babu, R.: Saliency unified: A deep architecture for simultaneous eye fixation prediction and salient object segmentation. In Proceedings of the IEEE Conference on Computer Vision and Pattern Recognition (pp. 5781-5790). (2016)

\bibitem{johnson2017study}
Johnson, S., Subha, T. D.: A Study on Eye Fixation Prediction and Salient Object Detection in Supervised Saliency. Materials Today: Proceedings, 4(2), 4169-4181. (2017)

\bibitem{tang2017prediction}
Tang, H., Chen, C., Bie, Y.: Prediction of human eye fixation by a single filter. Journal of Signal Processing Systems, 87(2), 197-202. (2017)

\bibitem{bylinskii2015saliency}
Bylinskii Z., Judd T., Borji A., Itti L., Durand F., Oliva A., Torralba A.: Mit saliency benchmark, http://saliency.mit.edu/. Last accessed 08 apr 2019

\bibitem{bylinskii2018different}
Bylinskii Z., Judd T., Oliva A., Torralba A., Durand F.: What do different evaluation metrics tell us about saliency models?. IEEE transactions on pattern analysis and machine intelligence (2018)

\bibitem{le2013methods}
Le Meur O., Baccino T.: Methods for comparing scanpaths and saliency maps: strengths and weaknesses. Behavior research methods, 45(1), 251-266. (2013)

\bibitem{wang2011simulating}
Wang, W., Chen, C., Wang, Y., Jiang, T., Fang, F., Yao, Y.: Simulating human saccadic scanpaths on natural images. In Computer Vision and Pattern Recognition (CVPR), 2011 IEEE Conference on (pp. 441-448). IEEE. (2011)

\bibitem{meur2017computational}
Meur, O. L., Coutrot, A., Liu, Z., Roch, A. L., Helo, A., Rama, P.: Computational model for predicting visual fixations from childhood to adulthood. arXiv preprint arXiv:1702.04657. (2017)

\bibitem{zheng2018task}
Zheng Q., Jiao J., Cao Y., Lau R. W.: Task-driven webpage saliency. In Computer Vision–ECCV 2018 (pp. 300-316). Springer, Cham. (2018)

\bibitem{yamada2010can}
Yamada K., Sugano Y., Okabe T., Sato Y., Sugimoto A., Hiraki, K.: Can saliency map models predict human egocentric visual attention?. In Asian Conference on Computer Vision (pp. 420-429). Springer, Berlin, Heidelberg. (2010)

\bibitem{yamada2011attention}
Yamada K., Sugano Y., Okabe T., Sato Y., Sugimoto A., Hiraki K.: Attention prediction in egocentric video using motion and visual saliency. In Pacific-Rim Symposium on Image and Video Technology (pp. 277-288). Springer, Berlin, Heidelberg.(2011, November)

\bibitem{li2013learning}
Li, Y., Fathi, A., Rehg, J. M.: Learning to predict gaze in egocentric video. In Proceedings of the IEEE International Conference on Computer Vision (pp. 3216-3223). (2013)

\bibitem{MLSmote}
Charte, F., Rivera, A. J., del Jesus, M. J., Herrera, F.: MLSMOTE: approaching imbalanced multilabel learning through synthetic instance generation. Knowledge-Based Systems, 89, 385-397. (2015)

\bibitem{sechidis2014information}
Sechidis K., Nikolaou N., Brown, G.: Information theoretic feature selection in multi-label data through composite likelihood. In Joint IAPR International Workshops on Statistical Techniques in Pattern Recognition (SPR) and Structural and Syntactic Pattern Recognition (SSPR) (pp. 143-152). Springer, Berlin, Heidelberg. (2014)

\bibitem{peng2005feature}
Peng, H., Long, F., Ding, C.: Feature selection based on mutual information criteria of max-dependency, max-relevance, and min-redundancy. IEEE Transactions on pattern analysis and machine intelligence, 27(8), 1226-1238. (2005)

\bibitem{jian2016multi}
Jian, L., Li, J., Shu, K., Liu, H.: Multi-Label Informed Feature Selection. In IJCAI (pp. 1627-1633). (2016)

\bibitem{nie2010efficient}
Nie F., Huang H., Cai X., Ding C. H.: Efficient and robust feature selection via joint $\boldsymbol\ell$2, 1-norms minimization. In Advances in neural information processing systems (pp. 1813-1821). (2010)

\bibitem{kimura2017mlc}
Kimura, K., Sun, L., Kudo, M.: MLC Toolbox: A MATLAB/OCTAVE Library for Multi-Label Classification. arXiv preprint arXiv:1704.02592 (2017)




\end{thebibliography}
\end{document}